# Optical properties of van der Waals heterostructure of uniaxially strained graphene on TMD

Partha Goswami

*D.B.College, University of Delhi, Kalkaji, New Delhi, India*

**Abstract** The spin and valley polarizations and plasmonics in Van der Waals heterostructures of strained graphene monolayer on 2D transition metal dichalcogenide (GrTMD) substrate are reported in this communication. The substrate induced interactions (SII) involve sub-lattice-resolved, and enhanced intrinsic spin-orbit couplings, the extrinsic Rashba spin-orbit coupling (RSOC), and the orbital gap related to the transfer of the electronic charge from graphene to the substrate. Furthermore, magnetic impurity atoms are deposited to the graphene surface and the corresponding exchange field is included in the band dispersion. A Rashba coupling dependent pseudo Zeeman term arising due to the interplay of SIIs was found to be responsible for the spin degeneracy lifting and the spin polarization. The latter turns out to be electrostatic doping and the exchange field tunable and inversely proportional to the square root of the carrier concentration. The strain field, on the other hand, brings about the valley polarization. The intra-band plasmon dispersion for the finite doping and the long wavelength limit has also been obtained. The dispersion involves the $q^{2/3}$ behavior and not the well known $q^{1/2}$ behavior. The uniform, uniaxial strain does not bring about any change in this behavior. However, the plasmon dispersion gets steeper for the wavevector perpendicular to the direction of strain and is flattened for wave vectors along the direction of the strain with the term responsible for the flattening proportional to the strain field. The stronger confinement capability of GrTMD Plasmon compared to that of standalone, doped graphene is an important outcome of the present work. One finds that whereas the intra-band absorbance of GrTMD is decreasing function of the frequency at a given strain field, it is an increasing function of the strain field at a given frequency.

**Main Text** Ever since the isolation and production of the graphene in 2004**[1,9],** the 2D materials**[10-11]** have attracted significant attention world-wide. Assembled from atomically thin layers of graphene, hexagonal boron nitride and other related materials, the van der Waals hetero-structures (vdWHs) **[12,13]**, have followed closely on the heels of the development of the 2D materials. The recent advances in vdWHs **[13-18]** have created a new touchstone in materials science unveiling unusual properties and new phenomena. These avant-garde structures not only offer a unique platform for the emerging devices with unprecedented functions, they also provide a fascinating dais for theoretical explorations through the handling of their confined electronic systems. For example, the higher degree of confinement and longer lifetimes of vdWH plasmons, accessed in graphene encapsulated boron nitride crystals **[19-21]**, have stimulated intense efforts to study such collective excitations triggered by the prospect of paving the way for architecting nano-photonic and nano-electronic devices and components. The bridging act of the whole spectral range from the mid-infrared to the terahertz (THz) band **[22]**, including the accommodation of the phonon modes, the ultra-confined, long-lived plasmon modes **[1– 9, 13-23]**, and the plasmon-phonon-polariton modes are new paradigm of such structures. The gate voltage and/or added impurities facilitate useful manipulation of these effects **[24,25]**. The vast possibilities presented by such hetero-structures suggest considerable future growth potential for this field in both fundamental studies and applied technologies.

The interesting and functional possibilities related to the optical and THz plasmons of strained graphene (Gr) monolayer on 2D transition metal dichalcogenide (TMD) substrate have been studied in this communication. The possibilities follow directly from the engineering of the enhanced spin-orbit coupling (SOC) in graphene through interfacial effects via coupling to the substrate. Apart from an orbital gap ($\Delta$)related to the transfer of the electronic charge from Gr to TMD, the substrate-induced interactions (SII) **[26]**are the sub-lattice-resolved, giant intrinsic SOCs due to the hybridization of the carbon orbitals with the d-orbitals of the transition metal( $\Delta_{soc}^{A}, \Delta_{soc}^{B}$) and the external electric field(($\boldsymbol{E}$) dependent extrinsic Rashba spin-orbit coupling (RSOC) ($\lambda_{\boldsymbol{R}}$ ($\boldsymbol{E}$)) that allows for the external tuning of the band gap in Gr-TMD and connects the nearest neighbors with spin-flip. These interactions are absent in isolated pristine, pure graphene monolayer. Furthermore, magnetic impurity (MI) atoms (such as Co) are deposited to the graphene surface. The MIs do not act as scatterer in our scheme; their effect is included in the band dispersion. We model the interaction between an impurity moment and the itinerant electrons in graphene with coupling term $J\sum_I \boldsymbol{S_I}.\boldsymbol{s_I}$, where $S_I$ is the I th-site impurity spin, $s_I = (1/2)\, a_{Is}^{\dagger} s_z a_{Is}$ or, $(1/2)\, b_{Is}^{\dagger} s_z b_{Is}$, $\boldsymbol{a_{Is}}$ ( $\boldsymbol{b_{Is}}$ ) is the fermion annihilation operator at site-I and spin-state s (=↑,↓) corresponding to the sub-lattice A(B), and $s_z$ is the z-component of the Pauli matrices. We make the approximation of treating the MI spins as classical vectors. The latter is valid for $|\boldsymbol{S}| > 1$. For a Co atom with the electronic structure as $4s^2 3d^7$, there are 7 electrons in d orbitals and d orbitals have 5 degenerate level of orbitals. There are three unpaired electrons with total spin angular momentum quantum number (3/2) and the total orbital quantum number (3). These values lead to total magnetic moment √27 Bohr magneton. Upon equating this with the formula for the spin quantum number only, the magnetic moment, viz. μ = √{4S(S+1)}, we obtain S → 2.15,–3.15 justifying the approximation made above. Upon absorbing the magnitude of the MI spin into the coupling constant $J$, and assuming that the impurities are located in the same sub-lattice, we obtain the dimensionless model interaction in momentum space as $\Delta\boldsymbol{H} = [M\sum_{k,s} sgn(s)\, (a^{\dagger}_{k,s}\, a_{k,s} + b^{\dagger}_{k,s}\, b_{k,s})]$ with $M = \frac{J|S|}{(\hbar v_F/a)}$ . The low-energy, dimensionless Hamiltonian for GrTMD around the Dirac points $\boldsymbol{K}$ and $\boldsymbol{K'}$ in the basis $(\boldsymbol{a_{k,\tau_z,\uparrow}}, \boldsymbol{b_{k,\tau_z,\uparrow}}, \boldsymbol{a_{k,\tau_z,\downarrow}}, \boldsymbol{b_{k,\tau_z,\downarrow}})$ where $\boldsymbol{a_{k,\tau_z,s}}$ ( $\boldsymbol{b_{k,\tau_z,s}}$ ) is the fermion annihilation operator for the momentum($k$)-valley($\boldsymbol{\tau}_z$) - spin ($s$) state corresponding to the sub-lattice A(B), may be written down as $H = [(\boldsymbol{\tau_z}\, \boldsymbol{a}k_x\sigma_x + \boldsymbol{a}k_y\sigma_y) + \Delta\, \sigma_z + M\, s_z + + \lambda_{\boldsymbol{R}}\, (\boldsymbol{E})\, (\boldsymbol{\tau}_z\sigma_x\, s_y - \sigma_y\, s_x) + (\boldsymbol{\tau_z}/2)\{\, \Delta^{\boldsymbol{A}}_{\boldsymbol{soc}}\sigma_{\boldsymbol{z}}(s_z + s_0) + \Delta^{\boldsymbol{B}}_{\boldsymbol{soc}}\sigma_{\boldsymbol{z}}(\boldsymbol{s}_z - s_0)\}]$. It may be noted that the Hamiltonian is time-reversal invariant in the absence of $M\, s_z$. The exchange field ($M$ ) can assist significantly the spin polarization due to the large shift of spin-up and down bands. We shall now consider the effect of strain field '$u_{ij}$' given by the well known **[27,28]** vector potential $\boldsymbol{A} = (\ A_x\ , A_{y,}\ 0\ )$ where $A_x = b_1(u_{xx} - u_{yy})$, and $A_y = b_2(2u_{xy})$ . These are the terms included as the minimal coupling in a study of strain-induced transport**[29,30]**. If non-uniform, the terms correspond to fictitious magnetic fields coupling with opposite sign to the two valleys**[31,32].** The theoretical suggestion that Landau levels, associated with non-uniform strain, can form in graphene was first given by Novoselov et al.**[31]** several years ago. This was followed by many reports of elastic Landau levels (ELL) in graphene samples**[33,34]**. We, however, consider a displacement field of the form $u(\boldsymbol{r}) = (2\rho\, y,\, 2\rho x)$,so that $u_{ii} = 0$ and $u_{ij} = \rho$. This uniform strain allows us to ignore ELL effects, such as the wavefunctions for both valleys having support only on the same sub-lattice in a pseudo-magnetic field giving rise to interesting outcome e.g. the opening up of a superconducting gap in the zeroth pseudo-Landau level in the presence of an on-site pairing potential. As in studies on semiconductors, the parameters ($b_1$, $b_2$) are to be determined experimentally or from ab-initio calculations**[27]**.The matrices $\boldsymbol{\tau_i}$, $s_i$ and $\sigma_i$, respectively, denote the Pauli matrices associated with the valley degrees of freedom, the real spin and the pseudo-spin of the Dirac electronic states. We shall replace below $\boldsymbol{\tau_z}$ by its eigenvalue $\xi = \pm 1$ for $\boldsymbol{K}(\boldsymbol{K'})$ cone. Here the nearest neighbor hopping is parameterized by a hybridization $t$, and $\hbar v_F/a = (\sqrt{3}/2)t$. The terms present in the Hamiltonian are made dimensionless dividing by the energy term ($\hbar v_F/a$).

Upon considering all the four substrate-induced interaction terms, the exchange field, and the strain field, the energy eigenvalues $\boldsymbol{\varepsilon} \equiv \varepsilon\, (\boldsymbol{ak_x}, \boldsymbol{ak_y}, \boldsymbol{M}, \boldsymbol{\rho})$ of the Hamiltonian are given by the quartic $\varepsilon^4 - 2\,\varepsilon^2\, b - 4\,\varepsilon\, c + d = 0$ where

$$b_\xi \equiv b_\xi(\boldsymbol{ak_x}, \boldsymbol{ak_y}, \boldsymbol{M}, \boldsymbol{\rho}) = [(a_1^2 + a_2^2 + a_3^2 + a_4^2)/4 + \boldsymbol{4}\lambda_{\boldsymbol{R}}^{\boldsymbol{2}} + (\boldsymbol{a}k_x)^2 + (\boldsymbol{a}k'_y)^2],\ \boldsymbol{a}k'_y = \boldsymbol{a}k_y - 2\xi\, b_2\rho, \quad (1)$$

$$c_\xi \equiv c_\xi(M) = (|\Delta^A_{soc}| - \Delta^B_{soc})\ [2\Delta\{|\Delta^A_{soc}| + \Delta^B_{soc}\} + 4\,\lambda_{\boldsymbol{R}}^{\boldsymbol{2}} - 4M\Delta\boldsymbol{\xi}], \quad (2)$$

$$\boldsymbol{d}_\xi \equiv d_\xi(\boldsymbol{ak_x}, \boldsymbol{ak_y}, \boldsymbol{M}, \boldsymbol{\rho}) = [a_1\, a_2\, a_3 a_4 - ((\boldsymbol{a}k_x)^2 + (\boldsymbol{a}k'_y)^2)(a_1\, a_3 + a_2 a_4) + ((\boldsymbol{a}k_x)^2 + (\boldsymbol{a}k'_y)^2)^2$$

$$- \lambda_R^2 (1-\xi)^2 (a_1 a_4) - \lambda_R^2 (1+\xi)^2 (a_2 a_3)], \quad (3)$$

$$a_1 = \Delta + M + \xi \Delta_{soc}^A, \; a_2 = -\Delta + M - \xi \Delta_{soc}^B, \; a_3 = \Delta - M - \xi \Delta_{soc}^A, \; a_4 = -\Delta - M + \xi \Delta_{soc}^B. \quad (4)$$

Note that, if $u_{ii} \neq 0$, one would have $ak_x \rightarrow ak'_x = ak_x + \xi A_x$ as well. These results yield the band structure

$$E_{\xi,s,\sigma}(ak_x, ak_y, M, \rho) = [s\sqrt{(z_0(ak_x, ak_y, M, \rho)/2)}\lambda_R + \sigma\{(ak_x)^2 + (ak'_y)^2 + \lambda^2_{\xi,s}(ak_x, ak_y, M, \rho)\}^{1/2}], \quad (5)$$

$$\lambda_{\xi,s}(ak_x, ak_y, M, \rho) = \{\beta^2_\xi(ak_y, M, \rho) - z_{0\xi}(ak_x, ak_y, M, \rho)/2 + s\sqrt{(2c^2_\xi(M)/z_{0\xi}(ak_x, ak_y, M, \rho))}\}^{1/2}, \quad (6)$$

$$\beta_\xi(ak_y, M, \rho) = [(a_1^2 + a_2^2 + a_3^2 + a_4^2)/4 + 4\lambda_R^2 - 4\xi b_2 \rho (ak_y) + 4(b_2\rho)^2]^{1/2}. \quad (7)$$

$$z_{0\xi}(ak_x, ak_y, M, \rho) \approx b_\xi(ak_x, ak_y, M, \rho) + \sqrt{\{d_\xi(ak_x, ak_y, M, \rho)\}}. \quad (8)$$

The band structure consists of two spin-chiral conduction bands and two spin-chiral valence bands. Because of the spin-mixing driven by the Rashba coupling, the spin is no longer a good quantum number. Therefore, the resulting angular momentum eigenstates may be denoted by the spin- chirality index $s = \pm 1$. Here $\sigma = +(-)$ indicates the conduction (valence) band. The gapped, spin- valley split bands involve a RSOC-dependent pseudo Zeeman field ( the term $s\sqrt{(z_{0\xi}/2)}\,\lambda_R$ in Eq.(5) )which couples with different signs to states with ↑,↓ spins. This field, however, couples with same sign but different manner to the states in the valleys due to the strain field $\rho$. Therefore, this pseudo field is expected to generate the spin-valley polarization in the system. The ($\lambda_R$ ($E$), $M$ ) dependence of the field indicates that the polarization is exchange field and external electric field tunable. In the absence of the substrate-induced interactions(RSOI is present though) and the exchange interactions, the band structure reduces to the spin- valley resolved energy dispersion of the graphene , viz. $E_{\xi,s,\sigma}(a|\delta k|, M) = \sigma\,[\{4\lambda_R^2 + (ak_x)^2 + (ak'_y)^2\} + 2s\lambda_R\sqrt{\{4\lambda_R^2 + 2((ak_x)^2 + (ak'_y)^2)\}}]^{1/2}$. If the RSOI is absent as well, then the band-structure reduces to the spin-valley degenerate energy dispersion of the strained graphene: $\grave{e}_\sigma = \sigma\sqrt{\{((ak_x)^2 + (ak'_y)^2)\}}$. It is gratifying to note that all the complexities present in the band structure is woven around the dispersion of the pure graphene.

On account of the strong, intrinsic spin-orbit interaction (SOI), as proposed by Kane and Mele **[35]**, the system acts as a quantum spin Hall (QSH) insulator for $M = 0$. In the Kane-Mele model (GrTMD system falls under this category), the effect of SOI is to create a bulk band gap together with the 'avoided crossing' between conduction and valence bands with opposite spin. There is no parity exchange (for the graphene system, the relevant bands are all pi-bands with the same parity) and no band inversion is necessary. The no band inversion condition is unlike that in the Bernevig–Hughes–Zhang (BHZ) model **[36]**, where the SOI induces the inversion at the high symmetry points in the Brillouin zone signaling the change in the parity of the valence-band-edge state and the transition from trivial to non-trivial insulator. We shall now see that the generic feature of QSH state, viz. the 'avoided crossing' is present in our scheme. In Figure 1 plots of band energies for $WSe_2$ as a function of the momentum component $a\,k_x$ with $\xi = \pm 1$, $a\,k_y = 0.0001$, $M = 0$ and the strain field $h = 2\,b_2\rho = 0.0001$ are shown. The (c↓,v↑) band-pair with $\xi = +1$ and the (c↑, v↓ ) band pair with $\xi = -1$ are characterized by the spin-orbit interaction (SOI) led 'avoided crossing' at the momentum (0, 0.0001). The anti-crossing of bands with opposite spins at the momentum(0, 0.0001) and not at (0,0) is due to the fact that the uniform strain field considered above represents a shift in momentum space of the Dirac cone. Similarly, there would be the anti-crossing between a pair of bands (c↓,v↑)((c↑, v↓)) with opposite spin and $\xi = +1$(( $\xi = -1$)) at the momentum (0.0001, ±0.0001) (not shown). These features are replicated in graphene on all TMDs.

As we exchange couple ($M$) the graphene layer in GrTMD system to localized magnetic impurities (MIs), such as substitutional Co atoms, we break the time-reversal symmetry (TRS). This may lead to the accessibility of the quantum anomalous Hall (QAH) state **[37,38,39]**. One finds that the graphene on TMDs [see Figures 1 and 2] is gapped at all possible exchange field values. As the exchange field ($M$) increases, the band gap narrowing takes place followed by its recovery to an extent. The essential features of the bands are (i) opening of an orbital gap due to the effective staggered potential and the intrinsic SOI, (ii) spin splitting of the bands due to the Rashba spin-orbit coupling and the exchange coupling, and (iii) the band gap narrowing and widening due to the many-body effect and the Moss-Burstein effect **[40]** respectively. The latter is due to the enhanced exchange effect. The plots( see Fig.2(a) ) for the Dirac point $\mathbf{K}$ shows that as the exchange field increases in $WSe_2$ / $WS_2$, the relevant band gap between the spin-down conduction band and the spin-up valence band gets narrower followed by the gap recovery. For the Dirac point $\mathbf{K}'$,we find that there is Moss-Burstein (MB) shift only and no band narrowing. It follows that the exchange field could be used for the efficient tuning of the band gap in graphene on TMD. The shift due to the MB effect is usually observed due to the occupation of the higher energy levels in the conduction band from where the electron transition occurs instead of the conduction band minimum. On account of the MB effect, optical band gap is virtually shifted to high energies because of the high carrier density related band filling. This may occur with the elastic strain as well as could be seen in Figure 2(b) . One may note that the band gap narrowing and the Fermi velocity $v_F$ renormalization, both, in Dirac systems, are essentially many body effects. The observation of the gap narrowing in graphene on $WSe_2$ / $WS_2$, thus, supports the hypothesis of $v_F$ renormalization**[41]**.Furthermore, (i) the direct information on the gap narrowing and the $v_F$ renormalization in graphene can be obtained from photoemission, which is a potent probe of many body effects in solids, and,(ii) new mechanisms for achieving direct electric field control of ferromagnetism are highly desirable in the development of functional magnetic interfaces.

We shall now briefly touch upon the spin and valley polarization, and the electron mobility of GrTMD; the latter is to justify the choice of TMD as the substrate. The pseudo-Zeeman term of the spectrum in Eq. (5) comes into being due the presence of the term ( $4\,\varepsilon\, c$ ) in the quartic $\varepsilon^4 - 2\,\varepsilon^2\, b - 4\,\varepsilon\, c + d = 0$. Without the term $4\,\varepsilon\, c$ ( in which case the magnitude of the sub-lattice resolved SOIs needs to be equal), the spectrum reduces to a bi-quadratic (with no Zeeman term) rather than a quartic. The term mimics a real Zeeman field with opposite signs for the two physical spin states. Its non-triviality lies in the valley states , the strain field, and the exchange field dependence. The role of this Zeeman field albeit the Rashba SOI in the spin-polarization ($P_s$) context, could be understood in the following manner: Recalling that the polarization $P_s$ is defined in terms of the spin-dependent conductance $G_s$ as $P_s = (G{\downarrow} - G{\uparrow})/(G{\downarrow}+G{\uparrow})$, and the spin-dependent current density magnitude is $j_s = [(ev_F/A\pi) \sum_\xi \int_{unfilled\ k} ak\ d(ak)\ \partial(E_{\xi,s,\sigma=+1,k}(a|k|,M) / \partial(ak)]$ for an applied constant electric field which, in the Drude's picture, is proportional to the spin-dependent conductance, one may write $P_s \sim (\lambda_R/\sqrt{2})[\sqrt{(z_0(\xi = +1,M))}+\sqrt{(z_0(\xi = -1,M))}]/\mu'$. Here A is a characteristic area and a sum over states $k$ is understood as an integral over all one-particle states. The contributions to the conductance from two Dirac nodes could be obtained by the sum$\sum_\xi$. We have approximated here $z_0(a\delta k, \xi, M)$ by $z_0(0, \xi, M)$ and $\lambda_s(a\delta k, \xi, M)$ by $\lambda_s(0,\xi,M)$ in view of their mild dependence on the wavevector. Thereafter, we have transformed the momentum integral to an energy integral in the zero-temperature limit. Though, admittedly, the finite temperature limit would have been appropriate. We introduce the quantity $\mu' = \mu/(\hbar v_F/a)$ where $\mu$ is the dimensionless chemical potential of the fermion number. All states below $\boldsymbol{\mu}$ are occupied. In view of Eqs. (5) and (6) we obtain $j_s \approx (e\mu'\, v_F/2A\pi) \sum_\xi [\eta\mu' - s\lambda_R \sqrt{(2z_0(0, \xi, M))}]$ where $\eta > 1$ . We have put the strain field equal to zero. The role of Rashba SOI as the polarization-usherer could be easily understood now. The polarization ($P_s$) turns out to be electric field ($E$) tunable as $\lambda_R$ is a function of $E$ **[26]**. Since there is a general relation **[42]**between $\mu$ and $V_g$ for a graphene-insulator-gate structure, viz.$\mu \approx \varepsilon_a [(m^2 + 2eV_g/\varepsilon_a)^{1/2} - m]$ where $m$ is the dimension-less ideality factor and $\varepsilon_a$ is the characteristic energy scale, the tunability of $P_s$ by the electrostatic doping is assured. Now the relation between $\mu$ and the carrier density may be given by $\mu \approx \hbar\, v_F \sqrt{(\pi|\mathbf{n}|)}\, sgn(n)$ where $sgn(n) = \pm 1$ for the electron(hole) doping and '$n$' is the carrier concentration, we obtain $P \sim n^{-1/2}$. Note that $P_s$ has opposite signs for the electron and hole doping. A 2D plot of the spin-polarization ($P_s$) in arb.unit as a function of the dimensionless exchange field ($M$), for the strain field ($h$) equal to zero, has been shown in Figure 3(a). The figure shows that the spin-polarization is an increasing function of the strain field.Since we have $z_0(\xi,M) \approx \lambda_R^2\ [b_\xi(\xi,M)+\sqrt{d_\xi(\xi,M)}]$ from Eq.(8), where $(b_\xi,d_\xi)$ are given by Eqs. (1) and (3), the dependence of the spin-polarization on SII parameters, such as the intrinsic SOI, orbital gap, etc.,is also obvious. The valley polarization ($P_v$), on the other hand, is defined in terms of the valley-dependent conductance $G_v$ as $P_v = (G_K - G_{K'})/(G_K + G_{K'})$. Since we do not have a pseudo Zeeman term with opposite signs for the two physical valley states, the previous approach for $P_v$ is not suitable. We, however, may write in Drude's framework itself the current density $j \sim \int D(\varepsilon)(e^2\tau/m_{eff})d\varepsilon$ where $D(\varepsilon) \sim 2|\varepsilon| / \pi(\hbar v_F)^2$ is the density of states (DOS) , and thus the valley polarization formula turns out to be $P_v \sim \int d\boldsymbol{k}\{\sum_{s,\sigma} (D(k)^{(K)}/m_{eff}^{(K)} - D(k)^{(K')}/m_{eff}^{(K')}) / \int d\boldsymbol{k} \sum_{s,\sigma} (D(k)^{(K)}/m_{eff}^{(K)} + D(k)^{(K')}/m_{eff}^{(K')})\}$ where $1/m_{eff}^{\xi=1\ (\xi=-1)} = (1/\hbar^2)\ \partial^2 E_{\xi=1\ (\xi=-1),s,\sigma}(\boldsymbol{ak_x}, \boldsymbol{ak_y}, \boldsymbol{M}, \boldsymbol{\rho}) / \partial k^2$. A 2D plot of the valley-polarization ($P_v$) in arb. unit as a function of the dimensionless strain field ($h$) for the exchange field ($M$) equal to zero is shown in Figure 3. The polarization is found to increase with strain, attain a maximum value, and thereafter decrease. It may be mentioned that Fujita et al. **[43]** had reported earlier the possibility of producing the valley polarized current in graphene considering a device, comprising of uniform uniaxial strain, and an out-of-plane magnetic barrier configuration. Thus, apart from the dependence on the strain field, $P_v$ can also be tuned by the exchange field. As regards the electronic mobility $\mu_0 = e\, v_F\, \tau / \hbar k_F$ , the scattering time $\tau$ is given by **[44]** $\tau^{-1} = 2n_i\, v_F\, k_F\, v_0^2 / (1 + \pi\alpha c/\varepsilon_r v_F)^2$ where $v_0 = V_0 / (\hbar v_F/a)$, $V_0$ is the scattering potential for the short-range point defect, and $n_i$ is the concentration of the impurity center. Here $\alpha = \frac{e^2}{4\pi\varepsilon_0\hbar c} = \frac{1}{137}$ is the fine-structure constant. Since the Fermi momentum $ak_F = ak_F(\mu', M) = (1/4) \sum_{s,\xi}\sqrt{\{(\mu' - s\sqrt{(z_{0\xi}/2)}\,\lambda_R)^2 - |\lambda_{-s,\xi}(M)|^2\}}$, and $\mu \approx \varepsilon_a [(m^2 + 2eV_g/\varepsilon_a)^{1/2} - m]$ (see ref.**[42]**) where $V_g$ is gate voltage, we observe that our expression for the mobility $\mu_0 = e\, v_F\, \tau/\hbar k_F$ shows its dependence on the exchange field $M$ and the variation in the gate voltage $V_g$. The strength of the scattering potential for the point defect also has tremendous effect on the mobility. In fact, greater the strength of the potential, lower is the mobility. The inter-band scattering processes have been completely ignored in this derivation. To include the intra- and the inter-valley scattering processes, one may utilize a model for screened scattering centers (SSC)of the Gaussian shape, with the screening length ($L$)spanning the range varying smoothly on the scale of the lattice constant($a$). The local potentials ( $L \sim a$) , due to the lifting of the prohibition on the inter-valley scattering, allow us to go beyond the scope of the single-valley scattering problem. A higher electron mobility of 300 $m^2/(V{\cdot}s)$ for electron densities of order $10^{16} m^{-2}$ on the TMD is accessed in comparison with that (17 $m^2/(V{\cdot}s)$ for electron densities less than $10^{16} m^{-2}$) in graphene on h-BN substrate **[19-21, 45].** Therefore, TMD, indeed, is an appealing substrate for graphene devices.

The plasmons are defined as longitudinal in-phase oscillation of all the carriers driven by the self-consistent electric field generated by the local variation in charge density. This collective density oscillations of a doped graphene sheet (Dirac Plasmons) are distinctively different($n^{1/4}$ dependence) from that ($n^{1/2}$ dependence) of the conventional 2D electron gas plasmons with respect to the carrier density ($n$) dependence. The former as well as the latter ones exhibit $q^{1/2}$ dependence as is well-known**[46,47,48]**.The broad reviews on graphene plasmonics with particular emphasis on the excitations in epitaxial graphene and on the influence of the underlying substrate in the screening processes could be found in refs.**[13,22,49]**. The great interest**[1-25,45-52]** evinced by the material science community in recent years in the graphene plasmons is linked to the facts that (i) the propagation of this mode has been directly imaged in real space by utilizing scattering-type scanning near-field optical microscopy**[24,25]**, (ii) the graphene plasmon is highly tunable and shows strong energy confinement capability**[50]**, and (iii) the graphene plasmons strongly couple to molecular vibrations of the adsorbates, the polar phonons of the substrate, and so on, as they are very sensitive to the immediate environment **[51]**. It may be recalled that using Maxwell's equations with appropriate boundary conditions, the plasmon dispersion could be obtained in the non-retarded regime ($q \gg \omega/c$) by solving the equation $\varepsilon + i(q/2\omega\varepsilon_0)\sigma(q,\omega) = 0$ where the dielectric constant $\varepsilon = (\varepsilon_1 + \varepsilon_2)/2$, $\varepsilon_1$ and $\varepsilon_2$ are the dielectric constants above and below the graphene sheet, $\varepsilon_0$ is the vacuum permittivity, $\sigma(q,\omega)$ is the wave vector dependent optical conductivity, $q$ is a wave vector, and $\omega$ is the angular frequency of the incident monochromatic optical field. Because of the finite scattering rate, $q = q_1 + iq_2$ has to be a complex variable with $q_2 \neq \mathbf{0}$ for the above equation to be valid. The ratio $R = q_1/q_2 = [\,\boldsymbol{\partial} / \boldsymbol{\partial q_1}(q_1\, Im\sigma(q_1,\omega))]/ Re\sigma(q_1,\omega)$ , referred to as the inverse damping ratio, is a measure of how many oscillations the plasmon makes before it is damped out completely. In view of the fact that the speed of light $c$ in vacuum is 300 time higher than the Fermi velocity of graphene, one could ignore the retardation effect in the equation above. The plasmon dispersion, corresponding to THz frequencies, in the present Gr-TMD system for a finite chemical potential, however, has been obtained within the RPA by finding the zeros of the dielectric function including only the. intra-band transition. The optical frequencies related functions corresponding to the inter-band transitions to be discussed later. In RPA, one writes the polarization function $\chi(\boldsymbol{aq},\boldsymbol{\omega}') = \chi_1(\boldsymbol{aq},\boldsymbol{\omega}') + i\ \chi_2(\boldsymbol{aq},\boldsymbol{\omega}')$in the momentum space in the long-wavelength limit $\hbar\omega \ll 2\ \varepsilon_F$(Fermi energy) , or $\hbar\omega' \ll \mathbf{2}\ \boldsymbol{ak_F}$(Fermi mometum), as

$$\chi(\boldsymbol{aq},\boldsymbol{\omega}') = \sum_{\xi}\sum_{k,\,s,s',\,\sigma,\sigma'=\pm 1} F_{\sigma,\sigma'}(\boldsymbol{k},\boldsymbol{q}) \times [\frac{\boldsymbol{n_{s,\xi,\sigma}(ak-aq)} - \boldsymbol{n_{s',\xi,\sigma'}(ak)}}{\{\hbar\boldsymbol{\omega}' + \varepsilon_{\xi,s,\sigma}(\boldsymbol{a(k-q)}) - \varepsilon_{\xi,s',\sigma'}(\boldsymbol{ak},) + \boldsymbol{i\eta}\}}], \quad (9)$$

where $(\hbar\omega/\hbar v_F a^{-1}) = \hbar\omega'$. Here, $\varepsilon$ and $\psi$ are single-particle energies and wave funtions, and $n_{\xi,s,\sigma}(ak) = [exp(\beta(\varepsilon_{\xi,s,\sigma}(ak) - \mu')) + 1]^{-1}$ is occupation function for the band $\sigma = \pm 1$. In the THz frequency range, contrary to the visible range case, the Fermi energy, due to the electrostatic doping, is usually much larger than the photon energy. Thus, the long-wavelength limit condition is easily satisfied in this case. For the graphene dispersion $E^{\pm}(k) = \pm t\ |\phi_{\boldsymbol{k}}| - \mu$ with $\phi_{\boldsymbol{k}} = [1+2exp\ (i3ak_x/2)\ \cos\ (\sqrt{3}ak_y/2)\ ]$, the spin-degenerate overlap of wave functions $F_{\sigma,\sigma'}(\boldsymbol{k},\boldsymbol{q})$ assumes the form $F_{\sigma,\sigma'}(\boldsymbol{k},\boldsymbol{q}) = ½[1 + \sigma\sigma' \boldsymbol{Re}(\boldsymbol{exp}(\boldsymbol{iaq}_x)\frac{\phi_{k-q}\,\phi^*_k}{|\phi_{k-q}||\phi_k|})]$. Here $\mu' = \mu/\ (\hbar v_F/a)$ is the dimensionless chemical potential of the fermion number. In the long-wave length limit, the band structure in Eq.(1) yields $\varepsilon_{\xi,s,\sigma}(\boldsymbol{a(k\pm q)}) - \varepsilon_{\xi,s,\sigma}(\boldsymbol{ak}) \approx \sigma \left\{\frac{a^2(q^2 \pm 2q.k \mp 4\xi\, b_2\, \rho q_y)}{2\lambda_{-s,\xi}}\right\}$. We shall not consider the spin-flip transitions here for simplicity. The dynamical dielectric (Lindhard)function, which is expressed as $e_{\xi,s,\sigma}(\boldsymbol{aq},\boldsymbol{\omega}') = 1 - \frac{V(q)}{(\frac{\hbar v_F}{a})}\chi_{\xi,s,\sigma}(\boldsymbol{aq},\boldsymbol{\omega}')$ where $V(q) = (e^2/2\,\varepsilon_0\,\varepsilon_r\delta q)$ is the Fourier transform of the Coulomb potential in two dimensions, $V(r) = e^2/4\pi\varepsilon_0\varepsilon_r r$ , and $\varepsilon_0$ the vacuum permittivity and $\varepsilon_r$ is the relative permittivity of the surrounding medium. The Lindhard function does not take into account interactions between electrons beyond RPA, impurities, and phonons. Upon making the Taylor expansion of $\boldsymbol{n_{s,\xi,\sigma}(ak-aq)}$ in Eq.(9): $\boldsymbol{n_{s,\xi,\sigma}(ak-aq)} = \boldsymbol{n_{s,\xi,\sigma}(ak)} + \boldsymbol{(aq)}\frac{\partial\, n_{\xi,s,\sigma}(ak)}{\partial\mu'}\frac{\partial\varepsilon_{\xi,s,\sigma}}{\partial(ak)}$ and replacing the derivative $\frac{\partial\, n_{\xi,s,\sigma}(ak)}{\partial\mu'}$ by the delta function in the zero temperature limit, after a little algebra, we obtain the real part of the intra-band Lindhard function as $e(\boldsymbol{aq},\boldsymbol{\omega}') \approx 1 - \mathcal{A}\int \boldsymbol{dk}\sum_{s,\xi}\ (\lambda^{-2}_{-s,\xi})\ \delta(\mu' - \varepsilon_{\xi,s,\sigma}(\boldsymbol{ak})) \times (\frac{(a|k|)^3\,(aq')^2(\hbar\omega')^{-3}}{\{(ak)^2 + (\lambda^2_{-s,\xi})\}^{\frac{1}{2}}})$, where $\mathcal{A} = \boldsymbol{\pi}(\frac{\frac{e^2}{2\,a\varepsilon_0\varepsilon_r}}{(\frac{\hbar v_F}{a})})$, $(aq_{s,\xi}')^2 = \left\{(\boldsymbol{aq})^2 - \mathbf{4\xi}\, \boldsymbol{b_2}\,\boldsymbol{\rho}\frac{qq_y}{k_{0,s,\xi}}\right\}$, and $(\boldsymbol{ak_{0,s,\xi}}) = [-\lambda^2_{-s,\xi} + \left(\mu' - \boldsymbol{s}\sqrt{\left(\frac{z_0(M)}{2}\right)}\,\lambda_{\boldsymbol{R}}\right)^2]^{1/2}$ .The latter, i.e. $(\boldsymbol{ak_{0,s,\xi}})$, determines the Fermi momentum. One notices now that, in the long wavelength limit $\hbar\omega' \ll \mathbf{2}\ \boldsymbol{ak_F}$ , the equation Re $e(\boldsymbol{\omega},\boldsymbol{aq}) = 0$ is a simple cubic in $\hbar\boldsymbol{\omega}'$with the only real solution

$$(\hbar\boldsymbol{\omega}'_{\boldsymbol{p}}) = \mathcal{A}^{1/3}\{\sum_{s,\xi}\ (\boldsymbol{aq_{s,\xi}'})^2\ Q_{s,\xi}(\mu, T=0, M)\}^{1/3},\ Q_{s,\xi}(\mu, T=0, M) = \frac{\left[\left(\mu' - s\sqrt{\left(\frac{z_0}{2}\right)}\lambda_R\right)^2 - (\lambda_{-s,\xi})^2\right]^{3/2}}{|(\lambda_{-s,\xi})|^3}. \quad (10)$$

We find that upon including the full dispersion of graphene on TMD and ignoring the spin-flip mechanism completely, there is only one intra-band collective mode and it corresponds to charge plasmons. In view of the above the quantity $Q^{1/3}$ $(\mu, T=0, M)$ approximately corresponds to the Fermi momentum $k_F$ ( $\boldsymbol{\varepsilon_F} = \hbar\boldsymbol{v_F k_F}$). Therefore, an approximate dependence $(\hbar\boldsymbol{\omega}'_{\boldsymbol{p}}) \propto n^{1/2}$ is expected for strained graphene on TMD ( and not $n^{1/4}$). Furthermore, since the Fermi momentum $ak_F = ak_F(\mu',M) = (1/4)\sum_{s,\xi}\sqrt{\{(\mu' - s\sqrt{(z_{0\xi}/2)}\lambda_R)^2 - |\lambda_{-s,\xi}|^2\}}$, and $\mu \approx \varepsilon_a\,[(m^2 + 2eV_g/\varepsilon_a)^{1/2} - m]$ (see ref.**[42]**) where $V_g$ is gate voltage, we observe that our expression of $(\hbar\omega'_p)$ shows its dependence on the exchange field $M$ and the variation in the gate voltage $V_g$ through the dependence on $\mu$. Therefore, in the THz range, the Plasmon frequency is a function of the exchange field and the gate voltage. The dependence of the Plasmon frequency on wave vector is of the form $(\hbar\omega'_p) \propto (aq)^{\frac{2}{3}}$. (and not the well known $q^{1/2}$ behavior) for graphene on TMD. We notice that the (wavevector)$^{2/3}$character of the plasmon branch as well as its approximate $n^{1/2}$ dependence on the charge density are not changed by uniform, uniaxial strain. However, the plasmon dispersion gets steeper for the wavevector perpendicular to the direction of strain and is flattened for wavevectors along the direction of the strain with the term responsible for the flattening proportional to the strain field. For typical graphene-on-TMD samples, $\boldsymbol{\varepsilon_F}$is of the order of 0:5 eV or lower. This indeed fixes the application range of graphene intra-band plasmonics to the THz and mid-IR band (wavelengths ≥1 μm). At a finite temperature the solution $(\hbar\omega'_p)$ and the real part of the polarization function may be written as

$$\hbar\omega'_p = \mathcal{A}^{1/3}\{\textstyle\sum_{s,\xi}\left(aq_{s,\xi}'\right)^2\; Q_{s,\xi}(\mu,T)\}^{1/3},\quad \chi_1(q,\omega') = \textstyle\sum_{s,\xi}\chi_{1,s,\xi}(q,\omega'). \tag{11}$$

Here $\beta = (k_BT)^{-1}$,

$$\chi_{1,s,\xi}(q,\omega') = \pi\left(\frac{aq_{s,\xi}'}{\hbar\omega'}\right)^3 Q_{s,\xi}(\mu,T),\; Q_{s,\xi}(\mu,T) = \int d(a\delta k)\frac{\beta(ak)^3}{4|\lambda_{-s,\xi}|^3}\cosh^{-2}\frac{\beta}{2}\left(\varepsilon_{\xi,s}(ak) - \mu'\right). \tag{12}$$

The imaginary part, on the other hand, is given by

$$\chi_2(\omega') \sim -\pi\textstyle\sum_{\delta k,\,s,\xi}[n_{s,\xi}(ak - aq) - n_{s,\xi}(ak)]\;\delta(\hbar\omega' + \varepsilon_{\xi,s,\sigma}(ak - aq) - \varepsilon_{\xi,s,\sigma}(ak))\;F(k,q). \tag{13}$$

Using a representation of the Dirac delta function, viz. $\delta(x) = Lim_{è\to 0}\,(1/2è\,\boldsymbol{\cosh^2\frac{x}{è}})$, Eq.(12) could be reduced to (10) in the zero-temperature limit. It may be mentioned that the most general expression of the dynamical polarization at finite temperature, chemical potential, impurity rate, quasi-particle gap, and magnetic field was presented by Pyatkovskiy and Gusynin **[52]** several years ago.

The effect of uniaxial strain on graphene's optical conductivity and absorbance have been described by introducing strain dependent hopping parameters into the standard tight binding Hamiltonian **[53,54]**. This deforms the Fermi surface into an ellipse and defines a fast and a slow optical axis, with the latter oriented closely along the direction of strain. We, however, follow a different approach to calculate the (intra-band) optical conductivity. As the first step, we note that the use of Re $e(\boldsymbol{\omega}, \boldsymbol{aq}) = 0$ above is essentially same as using the (plasmon dispersion) equation $\varepsilon_r + (i\,e^2/(2a\hbar\omega\varepsilon_0))\sum_{s,\xi}\left(aq_{s,\xi}'\right)\sigma_{\xi,s}(q,\omega) = 0$, where $\sigma_{\xi,s}(q,\omega)$ is the spin-valley dependent optical conductivity. To explain, one may replace $\sigma_{\xi,s}(q,\omega)$ by $\sigma_{\xi,s,RPA}(q,\omega) = i\,\sigma_0\left(aq_{s,\xi}'\right)^{-2}(\hbar\omega'/\pi)\chi_{s,\xi}(q,\omega)$ where $\sigma_0 = (e^2/4\hbar)$, $\hbar\omega' = \hbar\omega/\hbar v_F a^{-1}$, and $\chi_{s,\xi}(q,\omega) = \chi_{1,s,\xi}(q,\omega) + i\,\chi_{2,s,\xi}(q,\omega)$ is the dynamical polarization function. Therefore, $Im\sigma_{\xi,s,RPA}(q,\omega) = \sigma_0\left(aq_{s,\xi}'\right)^{-2}(\hbar\omega'/\pi)\chi_{1,s,\xi}(q,\omega) = \sigma_0\,(\hbar\omega')^{-2}\left(aq_{s,\xi}'\right)Q_{s,\xi}(\mu,T)$. Upon substituting this in the plasmon dispersion equation above , we get back Eq.(10). This is a good consistency check. Thus, the imaginary part of the optical conductivity $Im\sigma(q,\omega) = \sigma_0\,(\hbar\omega')^{-2}\sum_{s,\xi}\left(aq_{s,\xi}'\right)Q_{s,\xi}(\mu,T)$. The next step is to either use the second Kramers-Kronig (KK) relation**[55]** which gives the real part when the imaginary part is given, or use Eq.(13) to obtain the former. The use of (13) yields the frequency dependent

$$Re\sigma(\omega) = \sigma_0\ (\hbar\omega')\sum_{s,\xi,k} \left(aq_{s,\xi}{}'\right)^{-2} F_{\sigma\sigma}\ [\boldsymbol{n_{s,\xi}}\ (\boldsymbol{ak} - \boldsymbol{aq}) - \boldsymbol{n_{s,\xi}}\ (\boldsymbol{ak})]\ \delta(\hbar\boldsymbol{\omega}' + \varepsilon_{\xi,s,\sigma}(\boldsymbol{k}-\boldsymbol{q}) - \varepsilon_{\xi,s,\sigma}\ (\boldsymbol{ak})). \quad (14)$$

Upon making the Taylor expansion of $\boldsymbol{n_{s,\xi,\sigma}}(\boldsymbol{ak} - \boldsymbol{aq})$ ,i.e. $\boldsymbol{n_{s,\xi,\sigma}}(\boldsymbol{ak} - \boldsymbol{aq}) = \boldsymbol{n_{s,\xi,\sigma}}(\boldsymbol{ak}) + (\boldsymbol{aq})\ \frac{\partial\, n_{\xi,s,\sigma}(ak)}{\partial\mu'}\ \frac{\partial\varepsilon_{\xi,s,\sigma}}{\partial(ak)}$, we find that, in the long wavelength limit $\hbar\omega' \ll 2\,\boldsymbol{ak_F}$ , the real part of the optical conductivity is given by

$$Re\sigma(\omega') = \sigma_0 \sum_{s,\xi} \left((\boldsymbol{ak_F})^2/|\lambda_{-s,\xi}|\right)^2 \left(1 - \frac{|\lambda_{-s,\xi}|\,\hbar\omega'}{(\boldsymbol{ak_F})^2}\right)\left\{\mathbf{1} - \frac{4\xi\, b_2\,\rho}{ak_{0,s,\xi}}\right\}^{-1} (aq_{F,s,\xi}), \quad (15)$$

where $aq_{F,s,\xi} = \left\{\frac{\partial\left(n_{\xi,s,\sigma}(ak)\right)}{\partial\mu'}\ \boldsymbol{F}_{\sigma,\sigma}\right\}$ is the summand of the Thomas- Fermi screening wave vector of the system, viz. $aq_F = (1/4) \sum_{s,\xi} aq_{F,s,\xi}$ . The real part of the intra-band conductivity is, thus, decreasing function of frequency (in the long wave length limit).In the absence of strain, one finds $Re\sigma^{intraband}(\boldsymbol{\omega}') \sim \sigma_0\ (a\ q)^{-2}\ \hbar\omega' H(\boldsymbol{aq},\mu,\boldsymbol{\omega}')^{-1}$ where $(\hbar\omega/\ \hbar v_F a^{-1}) = \hbar\omega'$, and the function $H(\boldsymbol{aq},\mu,\boldsymbol{\omega}')$ is the height of the spectral function for the Gr-TMD (intra-band) plasmon case resembling a Lorentzian. The spectral function is symmetrical about the position of its maximum. This function characterizes the probability of electrons to undergo surface excitation in the surface region.The system under consideration, is the GrTMD vdWH. One may note that for the transmission of the THz and optical frequencies through the system, only the real part of the polarization function and optical conductivity are relevant. The reason for this is that the thickness of the vdWH in question is several orders of magnitude smaller than an radiation wavelength. In terms of $\sigma(\omega')$ given by Eq.(15), the optical absorbance $A(\omega')$ of the present system may be given by **[56,57]**

$$A(\omega') = (\pi\alpha\sigma(\omega')/\sigma_0)/\ [1 + (\pi\alpha\sigma(\omega')/2\sigma_0)]^2 = \sum_{s,\xi} \left(\mathbf{4}\pi\alpha\ \mathcal{C}_1\ /\ (2 + \pi\alpha\mathcal{C}_1)^2\right)$$

$$\times\ [1 - \{(2 - \pi\alpha\ \mathcal{C}_1)\ \varsigma\ \hbar\boldsymbol{\omega}'/\ (2 + \pi\alpha\ \mathcal{C}_1)\}], \quad (16)$$

$\varsigma \sim \frac{|\lambda_{-s,\xi}|}{(\boldsymbol{ak_F})^2}$, $\boldsymbol{C_1} = \left(\frac{(\boldsymbol{ak_F})^2}{|\lambda_{-s,\xi}|}\right)^{\mathbf{2}}\left\{\mathbf{1} - \frac{4\xi\, b_2\,\rho}{ak_{0,s,\xi}}\right\}^{-1} (aq_{F,s,\xi})$, $ak_F(\mu') = (1/4)\sum_{s,\xi}\xi\sqrt{\{(\mu' - s\sqrt{(z_{0\xi}/2)}\,\lambda_R)^2 - |\lambda_{-s,\xi}|^2\}}$, $\alpha = e^2\ /\ 4\pi\varepsilon_0\ c$ is the fine structure constant and $c$ is the speed of light. We find from the Figure 4 that whereas the intra-band absorbance of GrTMD $A\ (\omega)$ is decreasing function the frequency $\omega$ at a given strain field, it is an increasing function of the strain field at a given frequency. Unlike in standard semiconductors where the carrier type is fixed by chemical doping during the growth process, the Fermi level in graphene can be continuously driven (tuned)between the valence and conduction bands simply by the gate voltage **[58]**. Once again, since the Fermi momentum $ak_F = ak_F(\mu')$ and $\mu \approx \varepsilon_a\ [(m^2 + 2eV_g/\varepsilon_a)^{1/2} - m]$ (see ref.**[42]**) where $V_g$ is gate voltage, we observe that our expression of $\sigma(\omega)$ shows its dependence relative to the variation in the gate voltage $V_g$ through the dependence on $\mu$ . Therefore, in the THz range, the absorbance and the transmittance (which depend on the optical conductivity) are functions of the frequency, strain field, as well as the gate voltage. This property is expected to open up a wealth of interesting applications in THz photonic technologies. For pure graphene intra-band transitions, on the other hand, the corresponding expression**[59]** is $\sigma^{intraband}(\boldsymbol{\omega}') \sim \sigma_0\mathcal{C}(1 + i\,\hbar\ \omega'\tau')/\ (1 + (\hbar\ \omega'\tau')^2)$, where $\mathcal{C} = (4\tau'\boldsymbol{ak_F}/\pi)$ and $\tau'^{-1} = \tau^{-1}/\ \hbar v_F a^{-1}$ . One may notice that in the absence of scattering ($\tau'^{-1} << 1$), the expression corresponds to a purely imaginary quantity, signifying no dissipation of electric energy within Graphene. However, this is definitely unlikely to happen due to the inevitable phonon scattering and finite impurity doping. In the presence of scattering ($\tau'^{-1} >> 1$) this expression for the pure graphene intra-band conductivity gives $A(\omega') = (\mathbf{4}\pi\alpha\mathcal{C}\ /\ (2 + \pi\alpha\mathbf{C})^2) \times [\ 1 - \{(2 - \pi\alpha\mathcal{C})\ (\hbar\ \omega'\tau')^{\mathbf{2}}\ /\ (2 + \pi\alpha\boldsymbol{C})\}]$ **,**where $\alpha = e^2\ /\ 4\pi\varepsilon_0\ c$ is the fine structure constant and $c$ is the speed of light. Save for the fact that this is a quadratic (decreasing) function of the frequency $\omega$, there is form-wise similarity to the one obtained for GrTMD**.**

The results for Gr-TMD intra-band Plasmon are found to be rather remarkable. Since with hole doping the Fermi surface shifts to a lower energy, as a consequence the inter-band transitions with transition energy below twice the Fermi energy become forbidden. It leads to a decrease in higher frequency inter-band (optical) absorption. At the same time, the lower frequency (far infrared and terahertz (THz))free carrier absorption (i.e. intra-band transition) increases dramatically. Therefore, upon ignoring the spin-flip mechanism completely for simplicity, the intra-band transitions are needed to be considered only. Since, the focus is on the long wavelength regime, the transitions between two Dirac nodes located at different momentum could be neglected. These assumptions lead to a single collective mode for the present Gr-TMD system and it corresponds to the charge plasmons. One finds from the dispersion relation $(\hbar\omega/\hbar v_F a^{-1}) \sim (aq)^{2/3}$ or, $f \sim const.\times n^{1/2} q^{2/3}$ that the dimensionless quantity $r$, viz. the plasmon wavelength ($\lambda_{pl}$) and graphene lattice constant($a$) ratio, as a function of frequency ($f$) and the carrier concentration ($n$) is given by $r = K(n) f^{-3/2}$ where $K(n) \sim C n^{3/4}$, and $C$ is a constant. For $n \sim 10^{16}$ m$^{-2}$,and $a = 2.8\times 10^{-10}$m, we find $a\, K(n) \sim 10^{12}$ m-Hz$^{3/2}$. This leads to the plasmon wavelength $\lambda_{pl}$ as 1μ-m at THz and $10^{-3}$μ-m at the mid infrared spectral range. In comparison, for the standalone graphene sheet, the plasmon wavelength is $\lambda_0 = J(n) f^{-2}$ where $J(n) \sim C_1 n^{1/2}$; $C_1$ is a constant. For the same value of the carrier density one finds $J(n) \sim 10^{22}$ m-Hz$^2$. This leads to the plasmon wavelength $\lambda_0$ as 10 mm at THz and 1μ-m at the mid infrared spectral range. In the presence of scattering (the dimensionless relaxation time $\tau' = \tau/\hbar v_F a^{-1} = 0.1$) for $n \sim 10^{16}$ m$^{-2}$ and the inverse grating distance normalized by the Fermi wave vector ~0.1, on the other hand, the inverse damping ratio $R$ is found to be ~ 50 which corresponds to a reasonably well- defined Plasmon resonance. The stronger confinement capability of Gr-TMD plasmon is obvious from the ratio $\lambda_{pl} / \lambda_0$ . Unfortunately, to achieve an extreme plasmon confinement one has to sacrifice their propagation length( as $R\lambda_{pl} << R\lambda_0$ ). It is also limited by the inter-band absorption in the intrinsic samples and somewhat lower Drude absorption in the doped ones **[60]**.It is gratifying to see that a finite chemical potential or the gate voltage applied to a graphene sheet provides a conduction band for the electrons, allowing for plasmons supported by the graphene on TMD.

In conventional 2D electron gases (2DEGs), the Thomas-Fermi wave vector $\kappa$ is generally independent of the carrier density. However, for the pure graphene the screening wave vector is proportional to the square root of the density. Thus in pure graphene, the relative strength of screening ($\kappa/k_F$), where $k_F \sim \sqrt{(\pi n)}$ is the Fermi wave-vector, is constant. For Gr-TMD, a similar result is obtained in the context of the intra-band transitions. As observed by us, the relative strength of screening is nearly a constant with relative to the changes in the gate voltage (or, the carrier density) and the exchange field strength; at lower densities the behavior is slightly contrary to what one would expect. One is thus able to show that the characteristics linked to the screening in gated Gr-TMD is nearly insensitive to the substrate induced perturbations and the magnetic impurities. In the large momentum transfer regime, of course, the static screening increases linearly with wave vector due to the inter-band transition **[61]**. One also find that the stronger Rashba coupling (RSOC) has slight foiling effect on the Thomas-Fermi (TF) screening length ($\kappa^{-1}$). The RSOC parameter can be tuned by a transverse electric field and vertical strain. Interestingly, from the plots one also notice that the screening length is greater when the exchange field strength is greater.

As regards the optical absorbance ($A$) and the transmittance ($T$) due to the inter-band transitions, the latter at normal incidence is given by the equation $T(\omega) = (1+ \sigma_0^{interband}(\omega) / (2c\varepsilon_0))^{-2}$ where $\sigma_0^{interband}(\omega)$ is the real part of the inter-band optical conductivity. This can be obtained straightforwardly by solving the Maxwell equation with appropriate boundary conditions. Upon replacing $G_0$ by $\sigma_0=(e^2/4\hbar)$ is the frequency-independent universal sheet conductivity of the mass-less Dirac fermions above, one obtains $T(\omega) \approx 1-\pi\alpha$. The absorption, thus, corresponds to the well-known**[59 ]** value $\pi\alpha= 2.3\%$. The system under consideration, however, is the strained gapped graphene. In order to calculate $\sigma_0^{interband}(\omega)$ one now makes use of the relation **[62]**

$$\sigma_0^{interband}(\omega,V_g) =\sigma_0(4\pi/m_e^2\omega^2)\sum_{\sigma,\sigma'}\int (d^2k/(2\pi)^2 ( n_\sigma(k) - n_{\sigma'}(k))\, \delta(\hbar\omega+ E_\sigma(k)-E_{\sigma'}(k))F^{\alpha\beta}_{\sigma,\sigma'},$$

$$F^{\alpha\beta}_{\sigma,\sigma'}(k) = \prod^{\alpha}_{\sigma,\sigma'}(k) \prod^{\beta}_{\sigma,\sigma'}(k), \quad E_\sigma(k), E_{\sigma'}(k) < E_F \tag{17}$$

ignoring the many-body effects. Here, the indices $\sigma$ and $\sigma'$ denote the spin and all band quantum numbers for the occupied and empty states respectively, $k$ is the continuous quantum number related to the translational symmetry and restricted to the Brillouin zone, $E_F$ is the Fermi energy, $m_e$ denotes the electron mass, ω is the angular frequency of the electromagnetic radiation causing the transition, $\boldsymbol{n}_\sigma(\boldsymbol{k})$ is the Fermi-Dirac distribution function evaluated at energy $\boldsymbol{E}_\sigma(\boldsymbol{k})$,and $\prod^{\alpha}_{\sigma,\sigma'}(\boldsymbol{k}) = \langle\sigma',\boldsymbol{k}|\boldsymbol{p}_\alpha{}^{\boldsymbol{op}}|\sigma,\boldsymbol{k}\rangle$ is the transition matrix element of the α-component of the momentum operator $\boldsymbol{p}_\alpha{}^{\boldsymbol{op}}$ for a transition from the initial state $|\sigma, k\rangle$ with energy $\boldsymbol{E}_\sigma(\boldsymbol{k})$ into the final state $|\sigma', k\rangle$ with energy $\boldsymbol{E}_{\sigma'}(\boldsymbol{k})$. These matrix elements have been obtained from the band structure calculation above. In order to obtain an analytical form of (1), viz. $\sigma_0^{interband}(\omega,V_g) \approx \sigma_0\ (f(-\hbar\omega/2\ kT) - f(\hbar\omega/2\ kT))$where the Sokhotski-Plemelj identity $\{\omega - i\eta\}^{-1} = P(\omega^{-1}) + i\pi\delta(\omega) \approx i\pi\delta(\omega)$ has been used (and the imaginary part has been dropped) and $f$ stand for the Fermi-Dirac distribution, one applies the Dirac cone approximation. The approximation assumes that the conductivity is only contributed by the carriers on the Dirac cone. The Dirac cone approximation is only valid if the Fermi energy and the photon energy is within the visible range. Beyond that, the energy dispersion is not linear and one can no longer adopt the approximation. One finds $\sigma_0{}^{interband}(\omega)$ to be an increasing function of $\omega$ and almost independent relative to the variation in $V_g$ (see Figure 5). It is, in fact, (approximately) linearly varying with relative to $\omega$ in the limited, low photon energy range. The consistency of the linear relationship of $\sigma_0{}^{interband}$, with relative to $\omega$, with the Maxwell's law $\boldsymbol{\nabla}\times\boldsymbol{B} = i\omega\mu\varepsilon$ (1- $(i\sigma/\omega\mu_0\varepsilon_0 n^2)$) $\boldsymbol{E}$, where $\boldsymbol{B}$ and $\boldsymbol{E}$ are the magnetic and electric fields, respectively, $\mu$ ($\mu_0$) and $\varepsilon$ ($\varepsilon_0$) are permeability and permittivity, respectively, of the Dirac fermions (free space), $n = \sqrt{(\mu_r\varepsilon_r)}$ the optical index, and $\varepsilon_r$ is the complex relative permittivity, demands that the optical index ($n$) of the Gr-TMD, and, consequently, relative permittivity should be independent of frequency. In view of the pure graphene being known to possess dispersion-less optical index **[63,64]**, the observation that the calculated $\sigma_0{}^{interband}(\omega)$ is consistent with the Maxwell's law is not entirely un-founded. One may alternatively adopt the time-dependent first-order perturbation theoretic approach involving the Liouville-Von Neumann equation for the density matrix **[65]**, rather than use of the Kubo formula**[62]** as in here, to obtain the same result**[66]**. The imaginary part of the optical conductivity $\sigma_0{}^{interband}(\omega)$ is given by the first Kramers-Kronig (KK) relation**[55]**, $Im\sigma_0{}^{interband}(\omega) = (-2/\pi)\ \boldsymbol{P}\int_0^\infty \omega\ \boldsymbol{Re}\sigma_0^{\mathbf{interband}}(\varepsilon)\,(\varepsilon)\boldsymbol{d}\varepsilon/[\varepsilon^2 - \omega^2]$ where $P$ denotes the Cauchy principal value.

In conclusion, the Gr-TMD plasmons (optical as well as THz varieties) have unusual properties and offer promising prospects for plasmonic applications covering a wide frequency range. It has been also demonstrated here that the exchange field and the gate voltage can be used for efficient tuning of the band gap, the mobility, the intra-band plasmon frequency, and the optical conductivity. Whereas the spin-polarization is exchange field tunable, the valley polarization could be controlled by the strain field. We find that whereas the intra-band absorbance of GrTMD $A(\omega)$ is decreasing function the frequency $\omega$ at a given strain field, it is an increasing function of the strain field at a given frequency**.** We also notice that our major findings, viz.the (wavevector)$^{2/3}$character of the plasmon branch as well as its approximate $n^{1/2}$ dependence on the charge density, are not changed by uniform, uniaxial strain. However, the plasmon dispersion gets steeper for the wavevector perpendicular to the direction of strain and is flattened for wave vectors along the direction of the strain with the term responsible for the flattening proportional to the strain field. The stronger confinement capability of GrTMD Plasmon compared to that of standalone, doped graphene is an important outcome of the present work. The intra-band conductivity and absorbance can be controlled by the exchange field, too, As observed by us, the relative strength of screening is nearly a constant with relative to the changes in the gate voltage (or, the carrier density) and the exchange field strength. In the light of these findings we note that a direct, functional electric field control of magnetism at the nano-scale is needed for the effective demonstration of our results related to the exchange-field dependence. The magnetic multi-ferroics, like $BiFeO_3$ (BFO) have piqued the interest of the researchers world-wide with the promise of the coupling between the magnetic and electric order parameters.

**References**

[1] K S Novoselov, A. K. Geim, S. V. Morozov, D. Jiang,Y. Zhang, S. V. Dubonos, I. V. Grigorieva, and A. A. Firsov, Science 306 (5696), 666 (2004).
[2] A.H.C. Neto, F. Guinea, N.M.R. Peres, K.S. Novoselov, and A.K. Geim, Rev. Mod. Phys.81, 109 (2009).
[3] A. Woessner, M. B. Lundeberg, Y. Gao, A. Principi, P.Alonso-Gonzalez, M. Carrega, K.Watanabe,T. Taniguchi, G.Vignale, M. Polini, J. Hone, R. Hillenbrand, and F.H.L. Koppens, Nat. Mater. 14, 421 (2015).
[4] M. Gurram, S. Omar, S. Zihlmann, P. Makk,C. Schonenberger, and B. J. van Wees,Phys. Rev. B93, 115441 (2016).
[5] M. Gmitra, D. Kochan, P. Hogl, and J. Fabian, Phys. Rev. B 93, 155104 (2016).
[6] W. Yan, O. Txoperena, R. Llopis, H. Dery, L. E. Hueso,and F. Casanova,Nat Commun 7, 13372 (2016).
[7] P. Wei, S. Lee, F. Lemaitre, L. Pinel, D. Cutaia, W. Cha, F. Katmis, Y. Zhu, D. Heiman, J. Hone, J. S. Moodera,and C.-T. Chen, Nat. Mater.15, 711 (2016).
[8] Z. Wang, D.-K. Ki, J. Y. Khoo, D. Mauro, H. Berger, L. S. Levitov, and A. F. Morpurgo, Phys. Rev. X 6 , 041 020 (2016).
[9] Z. Wang, D.-K. Ki, H. Chen, H. Berger, A. H. MacDonald, and A. F. Morpurgo, Nat Commun 6, 8339 (2015)..
[10] E. C. T. OFarrell, A. Avsar, J. Y. Tan, G. Eda, and B.Ozyilmaz, Nano Letters 15, 5682 (2015).
[11]T. Georgiou, R. Jalil, B. D. Belle, L. Britnell, R. V. Gorbachev, S. V. Morozov, Y.-J. Kim, A. Ghonlinia, S. J.Haigh, O. Makarovsky, L. Eaves, L. A. Ponomarenko, A. K. Geim, K. S. Novoselov, and A. Mishchenko,Nat Nano 8, 100 (2013).
[12] A. K. Geim and I. V. Grigorieva, Nature 499, 419 (2013).
[13] W. Xia, L. Dai, P. Yu, X. Tong, W. Song, G. Zhang and Z. Wang , Nanoscale, **9**, 4324(2017).
[14] L. Viti, J. Hu, D. Coquillat, A. Politano, C. Consejo, W. Knap, M. S Vitiello, Adv. Mater. 28, 7390 (2016).
[15] C. R. Dean, A. F. Young, I. Meric, C. Lee, L.Wang, S. Sorgenfrei, K. Watanabe, T. Taniguchi, P. Kim, K. L. Shepard, and J. Hone, Nat. Nanotechnol. 5, 722 (2010).
[16] L. Britnell, R. M. Ribeiro, A. Eckmann, R.Jalil, B. D.Belle, A. Mishchenko, Y. J. Kim, R. V.Gorbachev, T. Georgiou, S. V.Morozov, A. N.Grigorenko, A. K. Geim, C. Casiraghi, A. H.Castro Neto, K.S.Novoselov,Science 340, 1311 (2013).
[17] A. C. Ferrari, F.Bonaccors, V.Fal'ko , K.S. Novoselov , S. Roche, P.Bøggild , S. Borini , F. H. L.Koppens, V.Palermo, N.Pugno, J.A.Garrido, R.Sordan , A.Bianco, L Ballerini, M.Prato, E.Lidorikis, J Kivioja, C.Marinelli, T Ryhänen, A Morpurgo, J.N.Coleman, V.Nicolosi, L.Colombo , A. Fert, M.Garcia-Hern-andez, A. Bachtold, G. F.Schneider, F.Guinea , C. Dekker, M Barbone, Z Sun, C Galiotis, A. N. Grigo-renkom, G.Konstantatos , A.Kis, M.Katsnelson , L Vandersypen, A .Loiseau, V.Morandi, D.Neumaier, E.Treossi, V.Pellegrini , M.Polini , A.Tredicucci , G. M.Williams, B.H.Hong , J. H.Ahn , J. M.Kim, H. Zirath, B. J.van Wees, H.van der Zant, L.Occhipinti, A.Di Matteo, I.A. Kinloch, T.Seyller, E.Quesnel , X. Feng, K.Teo, N. N.Rupesinghe, P.Hakonen, S. R.Neil, Q. Tannock, T.Löfwander, and .Kinaret,Nanoscale 7, 4598 (2015).
[18] W. J. Yu, Z. Li, H. Zhou, Y. Chen, Y. Wang, Y. Huang and X. Duan, Nat. Mater., 12(3), 246–252 (2013).
[19] A.S. Mayorov, R.V. Gorbachev, S.V. Morozov, L. Britnell,R. Jalil, L.A. Ponomarenko, P. Blake, K. S. Novoselov, K.Watanabe, T. Taniguchi, and A.K. Geim, Nano Lett. 11, 2396 (2011).
[20] L. Wang, I. Meric, P.Y. Huang, Q. Gao, Y. Gao, H. Tran,T. Taniguchi, K. Watanabe, L.M. Campos, D.A. Muller, J. Guo, P. Kim, J. Hone, K.L. Shepard, and C.R. Dean, Science 342 , 614 (2013).
[21] D. Bandurin, I. Torre, R.K. Kumar, M. Ben Shalom, A. Tomadin, A. Principi, G.H. Auton,E. Khestanova, K.S. Novoselov, I.V. Grigorieva, L.A. Ponomarenko, A.K. Geim, and M. Polini, Science 351, 1055 (2016).
[22] T.Low and P.Avouris, ACS Nano, 8, p. 1086-1101 (2014).
[23] P. Alonso-Gonz´alez, A.Y. Nikitin,Y. Gao, A. Woessner, M.B. Lundeberg, A. Principi, N. Forcellini, W. Yan,S. V elez, A.J. Huber, K. Watanabe, T. Taniguchi, F. Casanova, L.E. Hueso, M. Polini, J. Hone, F.H.L. Koppens, and R. Hillenbrand, arXiv:1601.05753 (Nature Nanotechnology, in press).
[24] Z. Fei, A.S. Rodin, G.O. Andreev, W. Bao, A.S. McLeod, M. Wagner, L.M. Zhang, Z. Zhao, M. Thiemens, G. Dominguez, M.M. Fogler, A.H. Castro Neto, C.N. Lau, F. Keilmann, and D.N. Basov, Nature 487 , 82 (2012).
[25] J. Chen, M. Badioli, P. Alonso-Gonz alez, S. Thongrattanasiri, F. Huth, J. Osmond, M. Spasenovi c, A. Centeno,A. Pesquera, P. Godignon, A. Zurutuza Elorza, N. Camara, F.J. Garc´ıa de Abajo, R. Hillenbrand, and F.H. L. Koppens, Nature 487, 77 (2012).
[26] M.Gmitra,D.Kochan, P.Hogl,and J.Fabian, arXiv: 1510. 00166 (2015);arXiv:1506.08954(2015); M. Gmitra, and J. Fabian, Phys. Rev. B 92, 155403 (2015);T. Frank, P. Hogl, M. Gmitra, D. Kochan, and J. Fabian, arXiv:1707.02124.
[27] Zhen-Gang Zhu and Jamal Berakdar, Phys. Rev. B 84, 195460 (2011).
[28] B. Amorim, A. Cortijo, F. de Juan, A.G. Grushin, F. Guinea, A. Gutiérrez-Rubio, H. Ochoa, V. Parente, R. Roldán, P. San-Jose, J. Schiefele, M. Sturla, M.A.H. Vozmediano, Phys.Rept. 617 (2016).

[29] V. M. Pereira, and A. H. Castro Neto, Phys. Rev. Lett. 103, 046801 (2009).
[30] V. M. Pereira, A. H. Castro Neto, and N. M. R. Peres, Phys. Rev.B80, 045401 (2009).
[31] F. Guinea, A. K. Geim, M. I. Katsnelson, K. S. Novoselov, Phys. Rev. B 81 035408 (2010).
[32] SP Lee,D Nandi,F. Marsiglio, and J. Maciejko, Phys. Rev. B 95, 174517 (2017).
[33] N. N. Klimov, S. Jung, S. Zhu, T. Li, C. A. Wright, S. D. Solares, D. B. Newell, N. B. Zhitenev, and J. A. Stroscio, Science 336 (6088) 1557 (2012).
[34] L. Meng, W.-Y. He, H. Zheng, M. Liu, H. Yan, W. Yan, Z.-D. Chu, K. Bai, R.-F. Dou, Y. Zhang, Z. Liu, J.-C. Nie, and L. He, Phys. Rev. B 87 205405 (2013).
[35] C.L. Kane and E.J. Mele, Phys. Rev. Lett. 95, 146802 (2005); C.L. Kane and E.J. Mele, Phys. Rev. Lett. 95, 226801 (2005);L. Fu and C. L. Kane, Phys. Rev B.76 , 045302 (2007).
[36] B.A. Bernevig and S.C. Zhang, Phys. Rev. Lett. 96, 106802 (2006); B.A. Bernevig, T.L. Hughes, and S.C. Zhang, Science 314, 1757 (2006); X.-L. Qi and S.-C. Zhang, Rev. Mod. Phys.83 , 1057 (2011).
[37] Xufeng Kou, Shih-Ting Guo, Yabin Fan, Lei Pan, Murong Lang, Ying Jiang, Qiming Shao, Tianxiao Nie, Koichi Murata, Jianshi Tang, Yong Wang, Liang He, Ting-Kuo Lee, Wei-Li Lee, and Kang L. Wang, Phys. Rev. Lett.113,199901 (2014).
[38] A. J. Bestwick, E. J. Fox, Xufeng Kou, Lei Pan, Kang L. Wang, D. Goldhaber-Gordon, Phys. Rev. Lett. 114, 187201 (2015).
[39] S.Y. Zhou, G.-H. Gweon, A.V. Fedorov, P.N. First, W.A. de Heer, D.-H. Lee, F. Guinea, A.H. Castro Neto, and A. Lanzara, Nature Materials 6, 770-775 (2007).
[40] T.S.Moss, Proceedings of the Physical Society. Sec. B, 67 (10) 775 (1954); E.Burstein, Phys. Rev. 93(3), 632 (1954).
[41] D. Kochan, S.Irmer, M.Gmitra, and J. Fabian, Phys. Rev. Lett. 115, 196601 (2015).
[42] G. I. Zebrev, Proceedings of 26th International Conference on Microelectronics(MIEL), Nis, Serbia, 159 (2008); arXiv.org 1102.2348(2011).
[43]T. Ando, J. Phys. Soc. Jpn.75, 074716 (2006).
[44] T. Fujita,M. B. A. Jalil,and S. G. Tan, *Applied Physics Letters* 97.4: 043508(2010).
[45] H. Hirai, H. Tsuchiy, Y. Kamakura, N. Mori, and M. Ogawa, *Journal of Applied Physics* 116, 083703 (2014).
[45] S. Das Sarma and E. H. Hwang, Phys. Rev. Lett.102,206412 (2009).
[47] M. Jablan, H. Buljan, and M. Soljaˇci´c, Phys.Rev. B 80, 245435 (2009).
[48] L. Ju, B.Geng, J.Horng, C.Girit, M.C.Martin, Z.Hao, H. A .Bechtel, X.Liang,A.Zettl,Y.R.Shen, and F. Wang, Nature Nanotechnology 6 630 (2011).
[49] A. Politano and G. Chiarello,Nanoscale 6 10927 (2014).
[50] V W Brar, M. S. Jang, M.Sherrott, J. J.Lopez, and H. A.Atwater, Nano Lett 13:2541–7(2013)..
[51] Z Fang , S. Thongrattanasiri, A. Schlather, Z. Liu, L. Ma, Y. Wang, P. Ajayan, P. Nordlander, and N.J. Halas, ACS Nano;7:2388–95 (2013).
[52] P.K. Pyatkovskiy and V. P. Gusynin, Phys. Rev. B83, 075422 (2011).
[53] V. Pereira, A. Castro Neto, and N. Peres, Phys. Rev. B 80 (4) 045401 (2009).
[54] F. M. D. Pellegrino, G. G. N. Angilella, and R. Pucci, Phys. Rev. B 81 (3) 035411 (2010).
[55] L.D. Landau, and E.M. Lifschitz,, "Electrodynamics of Continuous Media" , vol. 8 of "Course of Theoretical Physics," (Oxford, 1960),First edition.
[56] K. Kechedzhi and S. Das Sarma, Phys. Rev. B 88, 085403 (2013).
[57] L. Novotny, B. Hecht, Principles of Nano-Optics, Cambridge University Press( 2012).
[58] E. M. Hajaj, O. Shtempluk, V. Kochetkov, A. Razin, and Y. E. Yaish, Phys. Rev. B 88, 045128 (2013).
[59] T. Stauber, N.M.R. Peres, and A.K. Geim, Phys. Rev. B **78**(8) , 085432 (2008).
[60] A. Principi, M. Carrega, M. B. Lundeberg, A. Woessner, F. H. L. Koppens, G. Vignale, and M. Polini, Phys. Rev. B 90 (16), 165408(2014).
[61] S. Samaddar, I. Yudhistira, S. Adam, H. Courtois, and C.B. Winkelmann, Phys. Rev. Lett. 116, 126804 (2016).
[62] R. Kubo, J. Phys. Soc. Jpn. 12, 570 (1957).
[63] G. Teo, H. Wang, Y. Wu, Z. Guo, J. Zhang, Z. Ni, and Z. Shen, J. Appl. Phys.103, 124302 (2008).
[64] D. R. Lenski and M. S. Fuhrer, J. Appl. Phys. 110, 013720(2011).
[65] H.-P. Breuer and F. Petruccione, Open Quantum Systems(Oxford University Press, Oxford, 2002).
[66] V.P.Gusynin, S.G. Sharapov, and J.P. Carbotte, New Journal of Physics **11**(9), 095013(2009).

E-mail for corresponding author: physicsgoswami@gmail.com

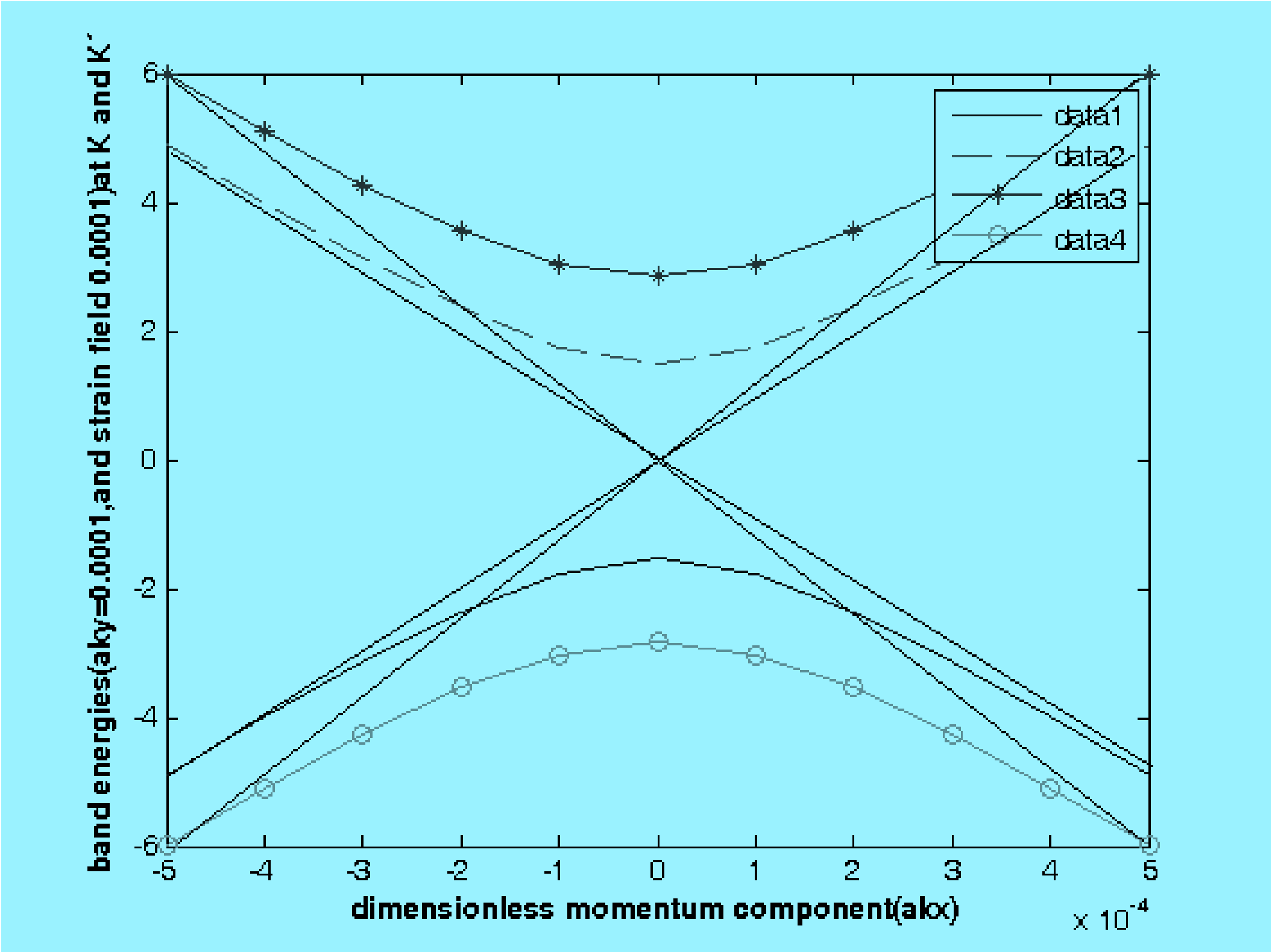

**Figure 1.** A plot of band energies for $WSe_2$ as a function of $a\,k_x$ with $\xi = \pm 1$, $a\,k_y = 0.0001$ and the strain field $2\,b_2\,\rho = 0.0001$characterized by the bulk band-gap between two pairs of bands with opposite spin, viz.(c↓,v↑) and (c↑, v↓ ) .There is avoided crossing at the momentum(0.0000, ± 0.0001) indicated by oblique intersecting lines for both the pairs of opposite spin projections. Here data1→ c↓, data2→ v↑, data3→ c ↑, and data4→ v↓.

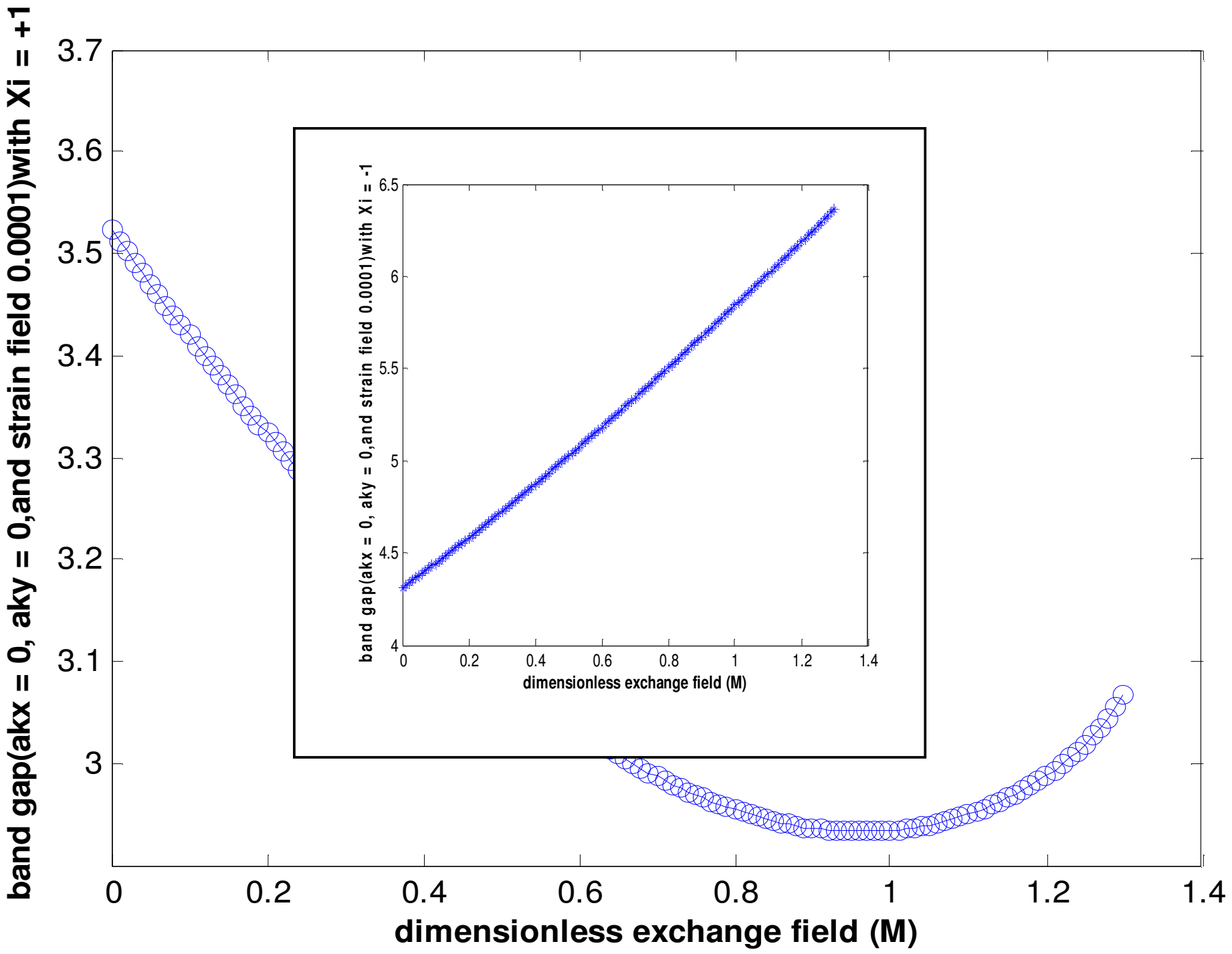

(a)

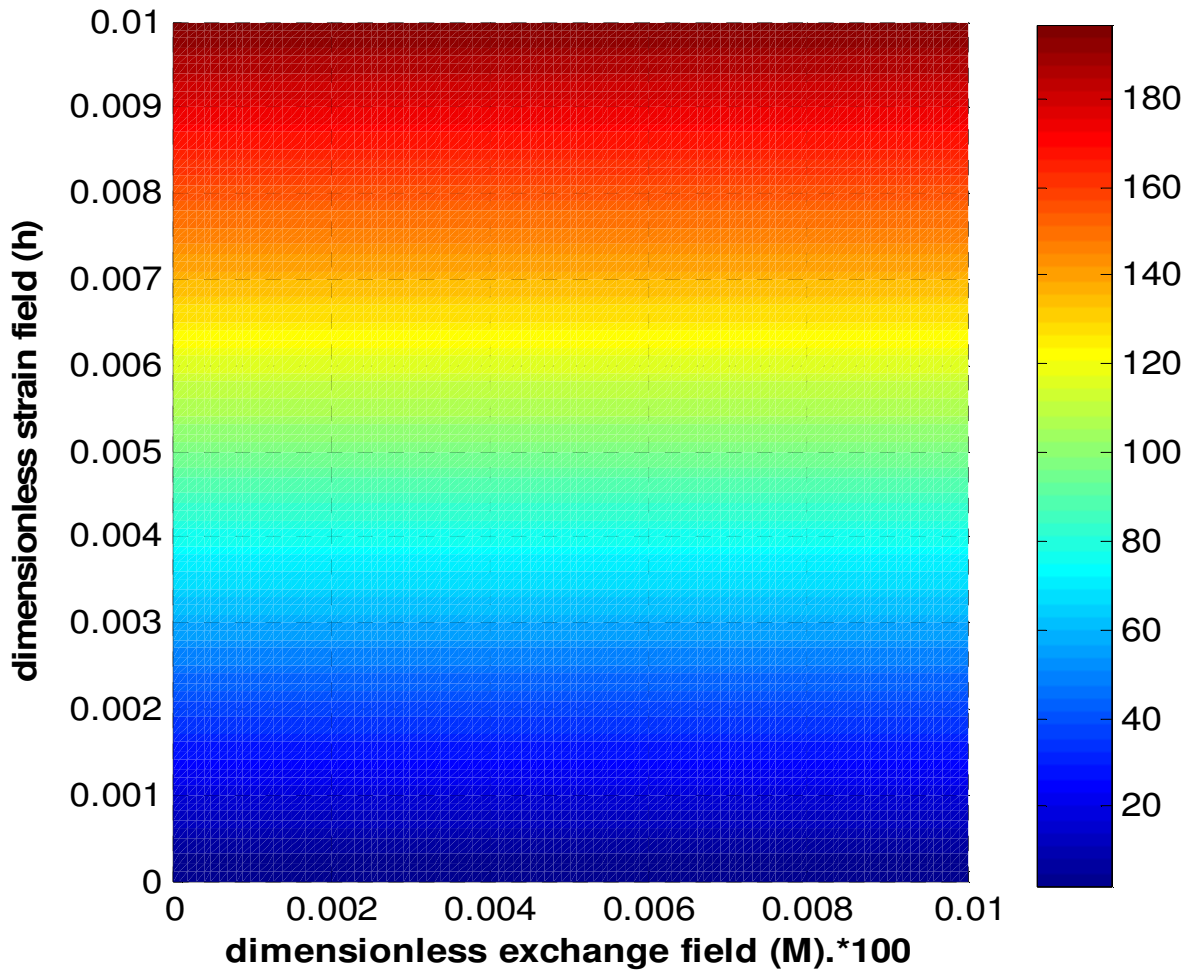

(b)

**Figure 2. (a)** A plot of the band gap with ξ = +1 between the two bands with opposite spin (c↓,v↑) in Figure 1for $WSe_2$ as a function of the dimensionless exchange field. The values of momenta $a\,k_x = 0 = a\,k_y$ , and the strain field $2\,b_2\,\rho = 0.0001$. There is decrease in the gap followed by the increase. For the remaining two opposite spin projections (c↑, v↓ ) with ξ = −1, however, there is increase as shown in the inset.(b) A contour plot of the band gap with ξ = +1 between the two bands with opposite spin (c↓,v↑) in Figure 1for $WSe_2$ as a function of the dimensionless exchange field and the dimensionless strain field. The gap increases considerably with the strain field but changes much less with the exchange field. The bulk band-gap with ξ = − 1 between two pairs of bands with opposite spin, viz. (c↑, v↓ ), will show the similar behavior.

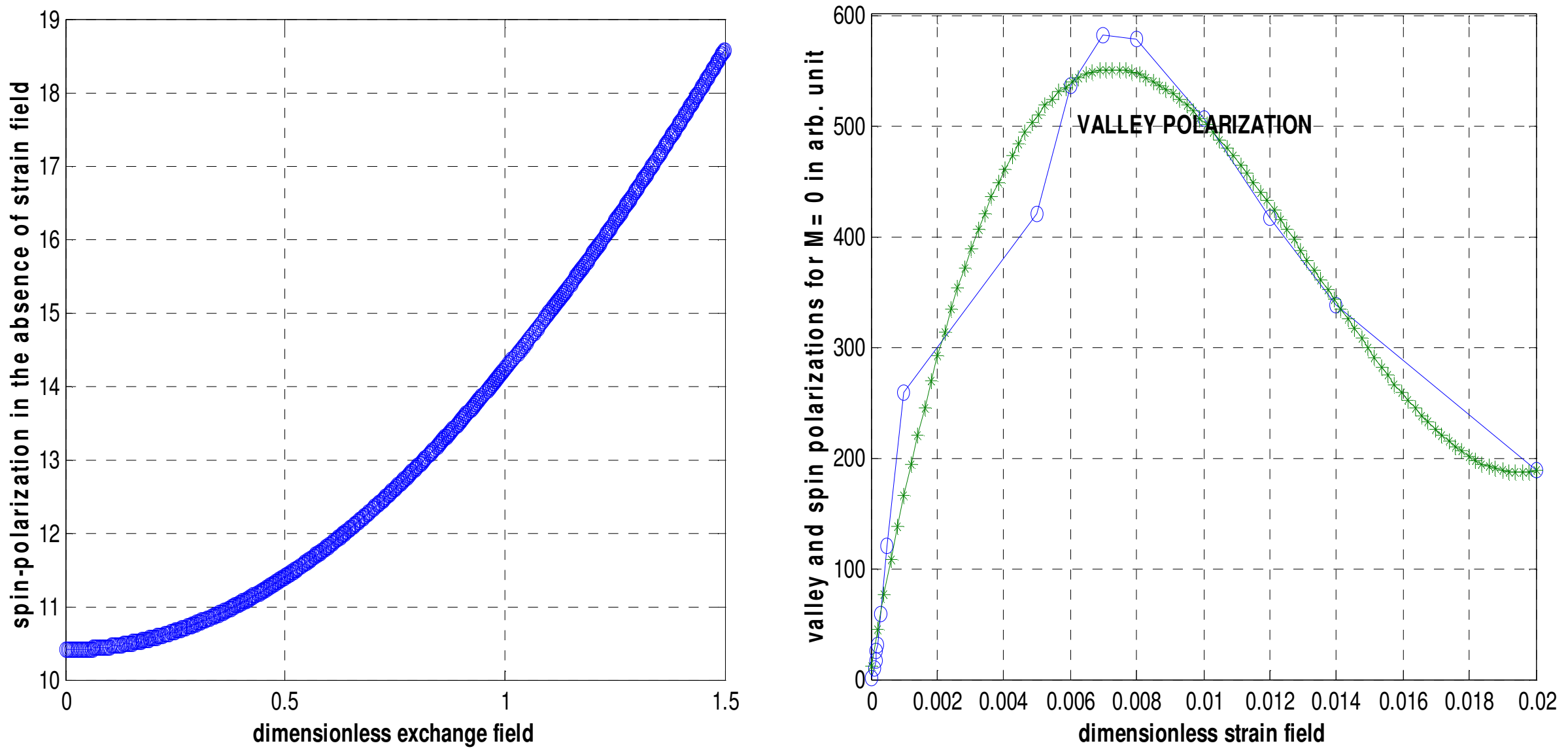

**(a)** **(b)**

**Figure 3.** (a) A 2D plot of the spin-polarization ($P_s$) in arb.unit as a function of the dimensionless exchange field ($M$) for the strain field ($h$) equal to zero. (b) A 2D plot of the valley-polarization ($P_v$) in arb.unit as a function of the dimensionless strain field ($h$) for the exchange field ($M$) equal to zero and the least square fit.

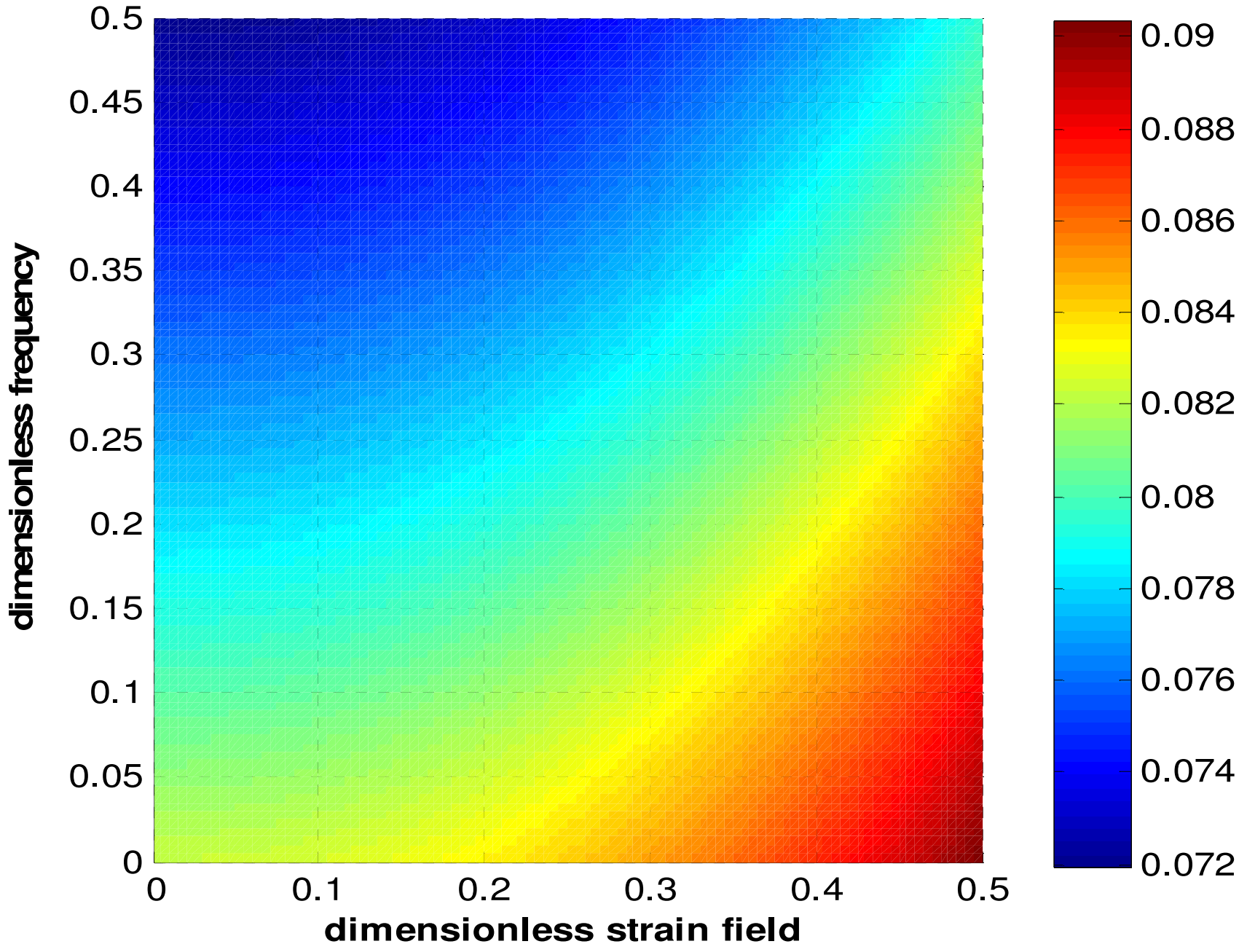

**Figure 4.** A contour plot of the absorbance A (ω) of graphene on TMD as a function of the dimensionless strain field and frequency. Whereas A (ω) is decreasing function of frequency at a given strain field, it is an increasing function of the strain field at a given frequency.

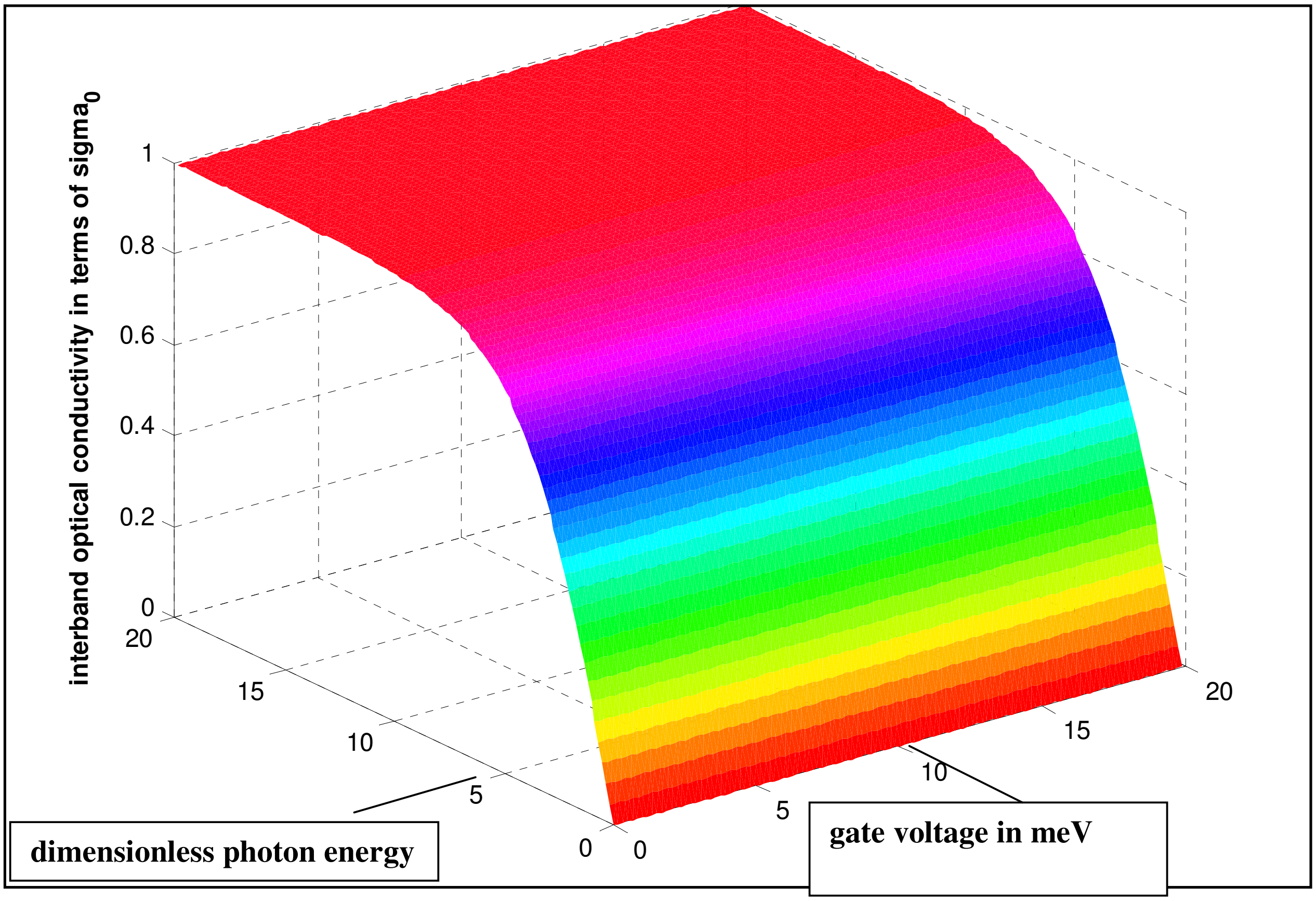

**Figure 5.** A 3D plot of the real part of the inter-band optical conductivity of graphene, in terms of $\sigma_0=(e^2/4\hbar)$, as a function of the dimensionless photon energy($\hbar\omega/kT$) and gate voltage in meV. Here the Boltzmann constant multiplied by the temperature( kT ) = 30 meV .